\documentclass[twocolumn]{jpsj2}
\usepackage{cuted,bm,times,mathptmx}
\newcommand{\kf}{k_{\mathrm{F}}}
\newcommand{\ut}{u_{\mathrm{t}}}
\title{
Possible Metastable State Triggered by Competition  \\ 
of Peierls State and Charge Ordered State
}

\author{Yukiko \textsc{Omori}, Masahisa \textsc{Tsuchiizu}, and 
 Yoshikazu \textsc{Suzumura}}

\inst{Department of Physics, Nagoya University, Nagoya 464-8602 }

\recdate{January 12, 2008; 
  published in J. Phys. Soc. Jpn.\ \textbf{76} (2007) 114709}

\abst{
 We examine a Peierls ground state
  and its competing metastable state
 in the one-dimensional quarter-filled Peierls-Hubbard model
  with  the nearest-neighbor repulsive interaction $V$
  and   the electron-phonon interaction ($\propto 1/K$ 
      with $K$ being the elastic constant).
From the mean-field approach,
we obtain the phase diagram for the ground state
  on the plane of parameters $V$ and $K$. 
The coexistent state of 
      the spin-density wave  and the charge ordering
    is realized for large $V$ and $K$.
With decreasing $K$, it 
   exhibits a first-order phase transition 
     to the unconventional Peierls state
   which is described by the 
    bond-centered charge-density-wave state.
 In the large region of 
   the  Peierls ground state in the phase diagram,
  there exists the metastable state
     where the energy takes a local minimum 
       with respect to the lattice distortion. 
  On the basis of the present calculation,  we discuss
   the photoinduced phase observed in the 
  (EDO-TTF)$_2$PF$_6$ compound.
 }

\kword{
 Peierls-Hubbard model, first-order phase transition,  charge ordering, 
 spin-density wave, quarter filling,  metastable state, 
 (EDO-TTF)$_2$PF$_6$ \\
}

\begin{document}
\maketitle

\section{Introduction}

 Electron correlation in quasi-one-dimensional molecular conductors  
 has been studied  extensively 
  \cite{Jerome,Gruner_rev} 
 where the spin density wave (SDW) state
   with the momentum 2$\kf$ ($\kf$ denotes the Fermi momentum)
    originates from the combined effect of repulsive interaction
   and the nesting of the Fermi surface.
 The former effect is relatively large in molecular conductors and 
   the latter effect becomes perfect for the one-dimensional band.  
  In particular, the quarter-filled band systems, which have been 
  the main subject in molecular conductors, 
      supply a rich variety of electronic states,
  \cite{Emery1979}
   e.g., 
   the charge ordering (CO) 
      \cite{Seo1997,Yoshioka2000}
  with a periodic array of charge disproportionation
 (+1, 0, +1, 0, $\cdots$).
   In addition to the pure 2$\kf$ SDW,
       the coexistent state  of the 2$\kf$ SDW and  2$\kf$
   charge-density wave (CDW)
       has been observed in the X-ray experiment on  
        (TMTSF)$_2$PF$_6$,\cite{Pouget,Kagoshima} 
   where  the coexistent state 
        has a  purely electronic origin
     due to the absence of lattice distortion.
 It  has been clarified theoretically that   
         the next-nearest repulsive interaction
         gives rise to such a 
          coexistence,\cite{Kobayashi1998,Tomio2000JPSJ} 
  and that
 the  state undergoes a first-order phase transition 
         into the normal state 
   with increasing temperature $T$.\cite{Tomio2001JPCS}

The molecular compound (EDO-TTF)$_2$PF$_6$,
 which is  another salt showing a typical quasi-one-dimensional system, 
   is a recent topic of the photoinduced phase transition.
\cite{Chollet2005,Ota2002,Drozdova2004,Onda2005}
  It has been reported that 
 the  first-order phase transition from metal to 
the Peierls insulator  
 occurs  at $T \simeq$ 278 K 
   and is followed by the fourfold periodicity of 
    the charge-rich sites and the charge-poor sites  
       with  the spatial variation of (+1, 0, 0, +1, $\cdots$).
  The large hysteresis is accompanied by the lattice distortion 
  indicating the strong coupling
   to the lattice through the electron-phonon (e-p)
  interaction.
This Peierls state 
  shows a distinctive feature of the $2\kf$ CDW
 where 
      the amplitude  of the lattice distortion takes a maximum 
        between the hole-rich sites of  the EDO-TTF molecules 
        (i.e., the bond ordering). 
  Although the spatial variation of the charge density 
  is similar to the conventional CDW,
  \cite{Pouget,Kagoshima,Kobayashi1998,Tomio2000JPSJ,Tomio2001JPCS} 
  the  spatial pattern of the lattice distortion found in 
   (EDO-TTF)$_2$PF$_6$ 
  has not been fully understood in the context of the previous theory.
   \cite{Kuwabara2003} 
 Furthermore, the conductivity of (EDO-TTF)$_2$PF$_6$ exhibits 
    a gigantic photoresponse,
   and the ultrafast insulator-to-metal transition is induced by the weak
   laser photoexcitation.
The mechanism of this behavior has been discussed by assuming the 
  existence of a metastable state without lattice distortion.
 However, it remains unclear if such a metastable state  can be 
    understood using the Peierls model in the presence of 
       the electronic correlation.

   In the present paper, we examine the possible origin of 
  the metastable state, which arises
   from the interplay of the Peierls state and the CO state.  
  In \S 2, the model with both the repulsive interactions and the e-p 
  interaction 
  is described and the formulation is given
   within the mean-field theory. 
    In \S 3 and \S 4, the ground state and the metastable state are 
     calculated      to obtain the phase diagram. 
 The possible parameter region for the metastable state
          is estimated.
  Section 5 is devoted to summary and discussions.

\section{Formulation}

We consider the 1/4-filled Peierls-Hubbard Hamiltonian given by
\cite{Su1979,Fukuyama1985} 
\begin{align}
  \label{eq:hamiltonian}
	H = & 
   - t 
   \sum_{j=1}^{N}
   \sum_{\sigma= \uparrow, \downarrow}
   (c_{j,\sigma}^\dagger c_{j+1,\sigma} + \mathrm{H.c.})
         \nonumber \\
     &    +  U\sum_j n_{j,\uparrow} n_{j,\downarrow}
         +  V\sum_j n_j n_{j+1}	
         \nonumber \\
     &  + t 
     \sum_{j\sigma}u_j(c_{j,\sigma}^\dagger c_{j+1,\sigma} + \mathrm{H.c.} )
                        \nonumber \\
      &
          + \frac{K}{2}\sum_j u_j^2  
          + \frac{\delta K}{2}\sum_{\mathrm{odd} \, j} u_j^2 ,  
\end{align}
where $c_{j,\sigma}$ (spin $\sigma = \uparrow, \downarrow$) 
 is the annihilation operator of the electron at the site $j$
  and $ n_j = \sum_{\sigma} c_{j,\sigma}^\dagger c_{j,\sigma}$.
We  impose the periodic boundary condition,
$c_{j+N, \sigma} = c_{j,\sigma}$. 
Quantities $U$ and $V$ denote the on-site and nearest-neighbor-site
   repulsive interactions, respectively. 
The term proportional to $u_j$ 
 (corresponding to the lattice distortion)  denotes 
the e-p interaction  and the last two terms represent 
the elastic energy with the elastic constant $K$
  and $K+\delta K$ for even and odd $j$, respectively. 
The alternating elastic constant given by $\delta K $ 
   plays a crucial role in the present paper. 
As for the lattice distortion $u_j$, we 
  take into account only the modulation with the momentum $2\kf(=\pi/2)$ 
  (corresponding to the lattice tetramerization), which 
  is relevant to the Peierls state in the (EDO-TTF)$_2$PF$_6$,
 and $u_j$ is rewritten as
 \begin{eqnarray} 
u_j =  \ut \cos \left( \frac{\pi}{2}j+\xi \right),
\label{eq:distortion}
\end{eqnarray}
where $\ut$ is the amplitude of the lattice tetramerization. 
 The elastic constant $K$ is scaled so as to include 
    the e-p coupling constant. 
In the present paper, we do not consider the possibility of 
  the lattice dimerization with the momentum $4k_F(=\pi)$. 
    \cite{Kuwabara2003,Clay} 
 Its detail is discussed in \S5.
    
After applying the Fourier transform,
 we define order parameters ($ m=0,1,2,3$) as 
\cite{Tomio2001JPCS} 
\begin{align}
   S_{mQ_0} =& \frac{1}{N} \sum_{\sigma=\uparrow,\downarrow}  
                \sum_{-\pi < k \leq \pi } \mathrm{sgn} (\sigma)  
                \left<c_{k,\sigma}^\dagger c_{k+mQ_0,\sigma}
                 \right>_{\mathrm{MF}} ,  
		                  \label{OPS} 
		                  \\
   D_{mQ_0} =& \frac{1}{N} \sum_{\sigma=\uparrow,\downarrow}  
                \sum_{-\pi < k \leq \pi }   
                \left<c_{k,\sigma}^\dagger c_{k+mQ_0,\sigma}
                \right>_{\mathrm{MF}}     
                   ,        \label{OPD}  
\end{align}
where
$Q_0 = 2\kf = \pi/2$,
 $S_0=0$, $D_0=1/2$, and 
\begin{align}
&  S_{Q_0}=S^*_{3Q_0}\equiv S_1 \mathrm{e}^{\mathrm{i} \theta} ,
\quad   D_{Q_0}=D^*_{3Q_0} \equiv  D_1 \mathrm{e}^{\mathrm{i} \theta'}  ,
    \nonumber \\
&  S_{2Q_0}=S^*_{2Q_0} \equiv S_2 ,
\quad  D_{2Q_0}=D^*_{2Q_0} \equiv  D_2 . 
\nonumber 
  \end{align} 
We note that
    $S_{1}$,  $S_{2}$,
      $D_{1}$, and   $D_{2}$ (which are real numbers)
         correspond to the amplitudes for the  2$\kf$ SDW, 4$\kf$ SDW,
           2$\kf$ CDW, and  4$\kf$ CDW, respectively,
      and $\theta$ and $\theta'$ are the phases for the  2$\kf$ SDW and 
            2$\kf$ CDW. 
It has been found that the relation $\theta' = \pi/2 + \theta$ holds  
             for $\ut = 0$.\cite{Tomio2000JPSJ}
In terms of these order parameters, 
 the mean-field Hamiltonian is written as 

\begin{strip}
\begin{align}
\label{eq:MF_Hamiltonian}
  H_{\mathrm{MF}}  =&
 \sum_{k,\sigma} 
\left\{
  \left( \epsilon_k+\frac{U}{4}+V \right) 
             c_{k,\sigma}^\dagger c_{k,\sigma}
 +
\left[
 \left( \Delta_{Q_0,\sigma}  
      + \frac{1}{2}  \ut {\mathrm e}^{\mathrm{i} \xi} 
 (1-\mathrm{i})(\cos k - \sin k) \right)
  c_{k+Q_0,\sigma}^\dagger c_{k,\sigma} + \mathrm{H.c.} \right]
+ 
    \Delta_{2Q_0,\sigma}
           c_{k,\sigma}^\dagger c_{k+2Q_0,\sigma} 
\right\}
\nonumber \\
 & 
- \frac{N}{4} \Biggl[ U \left(
         \frac{1}{4} + 2D_1^2- 2S_1^2 + D_2^2 - S_2^2 \right)
      + V ( 1-4D_2^2  ) {\Biggr]} 
 + \frac{N}{4}\left( K + \delta K \sin^2 \xi \right) u_{\mathrm t}^2 ,
\end{align}
\end{strip}

\noindent
 where	$\epsilon_k = -2t\cos k$ and  
\begin{align}
\Delta_{ Q_0,\sigma} 
&= \frac{U}{2}D_{Q_0}-{\mathrm{sgn}}(\sigma)\frac{U}{2}S_{Q_0},
\label{eq:S}\\
\label{eq:D}	                  
\Delta_{2Q_0,\sigma} 
&=
 \left( \frac{U}{2}-2V \right) D_{2Q_0}-{\mathrm{sgn}}(\sigma)\frac{U}{2}S_{2Q_0} .
\end{align}
The total  energy  is given by
   $ E(\ut, \xi) = \langle H_{\mathrm{MF}} \rangle_{\mathrm{MF}}/N$, 
 where $\langle \cdots \rangle_{\mathrm{MF}}$ is  
 the expectation value    over the mean-field Hamiltonian.
The ground-state energy is obtained by 
 minimizing the total energy 
 with respect to $\ut$ and $\xi$, 
 where the corresponding  equations are obtained as
\begin{align}
  & ( K  + \delta K \sin^2 \xi ) \ut
   \nonumber \\
  & = {\mathrm e}^{\mathrm{i} \xi} \frac{-1+\mathrm{i}}{N} \sum_{k,\sigma}  
                 (\cos k - \sin k) 
  \left<c_{k\sigma}^\dagger c_{k+Q_0,\sigma}
                 \right>_{\mathrm{MF}} 
                     + \mathrm{c.c.} ,  
\label{eq:ut}
\\
      &   -   \frac{\delta K}{2} \ut \sin (2\xi)  
   \nonumber \\  
   &  =  {\mathrm e}^{\mathrm{i} \xi} \frac{1+ \mathrm{i}}{N} \sum_{k,\sigma}  
                 (\cos k - \sin k) 
  \left<c_{k\sigma}^\dagger c_{k+Q_0,\sigma}  
                 \right>_{\mathrm{MF}}  + \mathrm{c.c.}  .
\label{eq:xi}
\end{align}
The parameters  $D_1, D_2, S_1, S_2$, $\ut$, $\theta$, $\theta'$, and $\xi$ 
 are evaluated  self consistently 
  using eqs. (\ref{OPS}), (\ref{OPD}), (\ref{eq:ut}), and
      (\ref{eq:xi}).
We also examine $E(\ut,\xi)$
  to investigate  the  possible  metastable state at $\ut=0$. 
The present calculation has been performed for $U=3$, 
where we take the hopping energy $t$ as unity.

\begin{figure}[b]
\centering
\includegraphics[width=8cm]{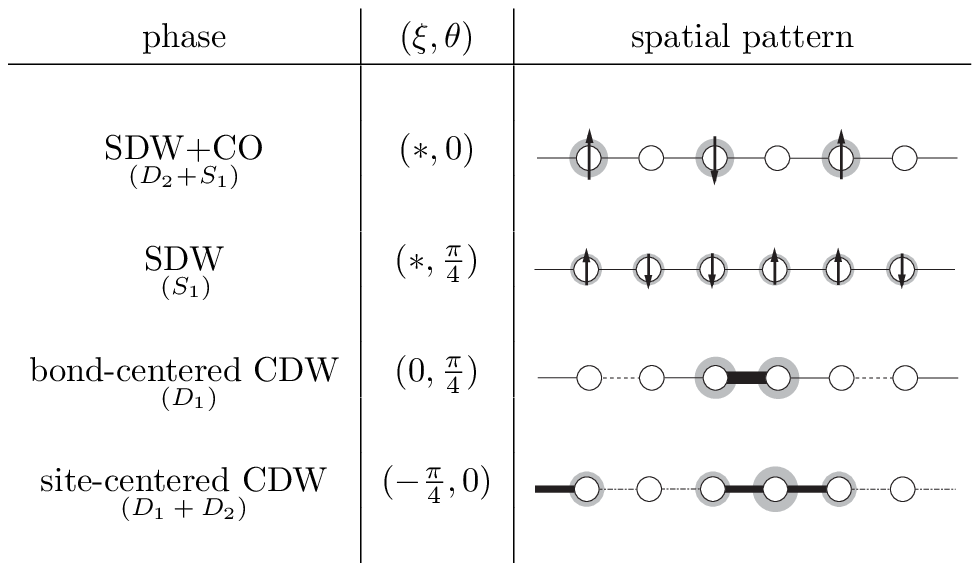}
\caption{ 
Several possible ordered states obtained in the present model.
The notation $(\xi,\theta)$ denotes the set of the phases 
 for the lattice tetramerization ($u_j=\ut\cos (\pi j/2 + \xi)$) 
 and  for  the order parameter of 
 the SDW ($S_1 \mathrm{e}^{\mathrm{i} \theta}$)
or the CDW ($D_1 \mathrm{e}^{\mathrm{i} (\pi/2 + \theta)}$). 
 The circle with the shaded region 
 denotes charge-rich sites and the bold line corresponds to the 
 bond with large transfer energy.   
}
\end{figure} 

  In Fig.~1, we present the states  obtained in our  Hamiltonian. 
  Two kinds of phases of $\xi$ and $\theta$ correspond to those of 
   order   parameters   for the lattice tetramerization  and the $2\kf$ SDW,
    respectively:
\begin{eqnarray}
\ut \cos \left(\frac{\pi}{2} j + \xi\right), \quad
 S_1 {\mathrm e}^{\mathrm{i} \theta} .
\nonumber 
 \end{eqnarray}   
The symbol $*$ in Fig. 1 implies the state with $\ut=0$. 
   The SDW + CO ($D_2 + S_1$) 
  is the charge-ordered state with the alternation 
 of the charge-rich/poor sites. 
 The pure SDW ($S_1$) has a spin amplitude, the maximum of 
   which  is located on the bonds 
  in order to gain the transfer energy.
 Both the SDW+CO state and the SDW state do not have 
   any lattice distortion, namely,  $\ut = 0$.     
The bond-centered CDW and site-centered CDW states are the Peierls
   states, i.e., the nonmagnetic states with finite lattice distortion
  (tetramerization) and with $2\kf$ charge modulations.
The lattice tetramerization takes the maximum amplitude
  on the bonds in the former state, 
  while it does so on the sites in the latter state.
We note that all these states exhibit the insulating behavior
   due to the presence of the $2\kf$ density wave ($S_1$ or $D_1$), which
  yields  the  gap in the dispersion at  the Fermi energy.

We note  the relevance of the present analysis 
 to the state of the quasi-one-dimensional (EDO-TTF)$_2$PF$_6$ compound.
 The Peierls state observed at low temperatures
   is  precisely of the bond-centered CDW type. \cite{Chollet2005}
Therefore, in the present paper, 
 such a type of the CDW is mainly examined 
  to find  the possible metastable state, which could  be 
  the origin of the photoinduced phase 
  realized  after releasing the lattice distortion.

\section{Effect of  Alternating Elastic Constant on Ground State}

In order to clarify the starting point of the present work,
 we begin to examine 
 the case for $\delta K = 0$, 
 in  which some results  can be compared with the previous works.  

First, we briefly recall the 
electronic state  in the absence of the e-p interaction.\cite{Seo1997}
There is a critical value $V_{\mathrm{c}} \simeq  0.28$
where   the state with $S_1 \not=0$ and $D_2 =0$ is obtained
  for $V < V_{\mathrm{c}}$, 
 and that with 
   $S_1 \not=0$ and $D_2 \not= 0$ is obtained for $V > V_{\mathrm{c}}$. 
 The CO state for  $V > V_{\mathrm{c}}$ has
  a charge disproportionation with two-fold periodicity (Fig. 1). 
 The phase transition  at $V=V_{\mathrm{c}}$ is of the first order where
  the both order parameters $D_2$ and $S_1$
   exhibit discontinuous change.
 The difference in SDW ($S_1$) between the case of $V<V_{\mathrm{c}}$ and 
 that of $V>V_{\mathrm{c}}$ 
 is that  the  maximum of the spin amplitude is on the bonds
  in the former case while 
  it is  on the sites in the latter case.
Such a difference  can be understood from 
 the commensurability energy for the $2\kf$ SDW 
  ($S_{Q_0} \equiv S_1 {\mathrm e}^{\mathrm{i}\theta}$) at 
   quarter filling,
 which decreases continuously 
 and vanishes only at $V=V_{\mathrm{c}}$.
Actually,  the $\theta$ dependence of 
  eq.~(\ref{eq:MF_Hamiltonian}) 
  takes the form
\cite{Suzumura1997}  
\begin{eqnarray}
\label{eq:Eg}
 E_g(\theta) = \mathrm{const.} + C(V) \, \cos 4 \theta ,
 \end{eqnarray} 
where 
  the $V$ dependence of 
    $C(V)$ is given by
  $C(V)= C_{00} - C_{01} V - C_{03} V^3 + \cdots $
  with $C_{0j}$ being  positive numbers.
Thus, one finds 
  $C(V) > 0$,  i.e., $\theta = \pi/4$ (bond-centered density wave), 
for  $V < V_{\mathrm{c}}$, while one finds 
  $C(V) <0 $,  i.e., $\theta = 0$ (site-centered density wave), 
  for $V > V_{\mathrm{c}}$.
We note that, due to such a sign change of $C(V)$,    
 the collective mode for the phase fluctuation exhibits 
  the vanishing of the commensurability gap at $V=V_{\mathrm{c}}$ 
 leading to the metallic property.
 In fact, a noticeable behavior emerges in the optical conductivity 
 $\sigma (\omega)$ ($\omega$ being the frequency),
   where the static conductivity $\sigma (0)$ becomes finite 
   at $V=V_{\mathrm{c}}$ even for the commensurate SDW state.
   \cite{Tomio2002JPSJ}  

Next, we consider the case with the e-p coupling but with $\delta K =0$.
In Fig.~2, we show the phase diagram of the ground states
  obtained from 
 eqs.~(\ref{OPS}), (\ref{OPD}), (\ref{eq:ut}),  and (\ref{eq:xi}).
  The bold line denotes the 
   boundary for the first-order phase transition,  
   which is estimated by comparing the minimum energy of $E(u_{\rm t},\xi)$.  
For large $K$ (i.e., for  small e-p coupling), 
we obtain  $\ut = 0$    
 and   there is a critical value $V_{\mathrm{c}} \simeq 0.28$,  where  
  the SDW state is obtained for $V < V_{\mathrm{c}}$ 
  and 
  the SDW+CO state is obtained for $V  > V_{\mathrm{c}}$. 
 With decreasing $K$, the state with finite lattice tetramerization
 ($\ut \neq 0$) appears. 
There is the $D_1 + S_1$ state 
 in between the SDW state and the bond-centered CDW  state,
where both boundaries do not depend on $V$. 
Such a $V$-independent result is ascribed to the 
 mean-field treatment in which   
 $V$ does not contribute to  the mean field 
  for $S_1$ and $D_1$ [see eqs.~(\ref{eq:S}) and (\ref{eq:D})].
For the intermediate $K$ and small $V$, 
  there exists  the region of the bond-centered CDW. 
 This state resembles the BCDW state suggested by Clay \textit{et al}. 
 (Fig.~3(b) and Fig.~4 in ref. \citen{Clay})
 and the DM+SP state suggested by Kuwabara \textit{et al}.
  (Fig.~2 in ref. \citen{Kuwabara2003}), 
  in the sense that the amplitude of the lattice tetramerization 
  takes a maximum on the bonds; i.e., bond-centered state.

\begin{figure}[t]
\centering
\includegraphics[width=8.5cm]{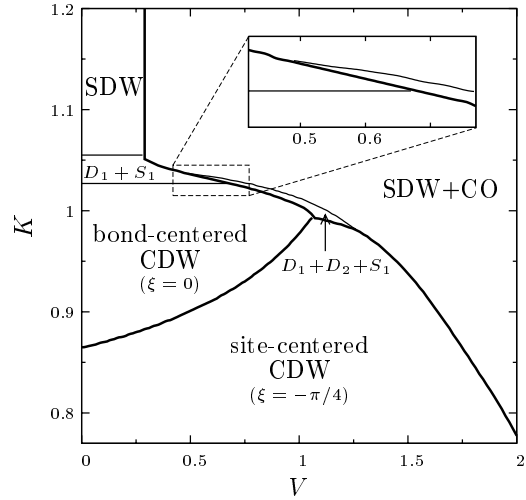}
\caption{
Phase diagram on the plane of $V$ and $K$ 
 for $\delta K = 0$ and $U=3$.
 The bold (thin) solid line  corresponds  to 
   the first- (second-)order phase transition.
The region framed by the dotted rectangle is enlarged to show
     clearly several kinds of phases. 
}
\end{figure}

Here, we note the effects of the quantum fluctuation 
  and the lattice dimerization, which are not taken into account  
     in  the  present mean-field treatment. 
 It is possible that the   magnetic state of the SDW+CO 
   turns  into a nonmagnetic state
     in the presence of the quantum fluctuation and  the e-p coupling. 
 Compared with the case in  ref. \citen{Kuwabara2003},
  the mean field stabilizes  the CO and the charge disproportionation 
  in which the  strong correlation and the quantum fluctuations 
   take important roles.   
   Thus, in the  mean-field treatment, 
  the boundary between the bond-centered CDW state and 
  the site-centered CDW state 
    is shifted to the weak coupling region,  
  and  the SDW+CO state is stabilized for relatively small $V$. 
 It is also expected  that, due the  quantum effect,  
 the boundary between the SDW+CO and site-centered CDW states 
   becomes a crossover 
    and the nonmagnetic state would be realized in both regions. 
Furthermore the  lattice dimerization 
   can be expected to coexist with the bond-centered CDW state.
   \cite{Clay,Kuwabara2003}
In the sense of the variational principle,
 the region for the bond-centered CDW is extended 
   by introducing the lattice dimerization.

The bond-centered CDW state 
  originates from the facts that
   the on-site repulsive  interaction   $U$   separates  two electrons 
     being on the same site, and that the bond-centered ordering gives rise to 
     the large gain of the transfer energy.  
  For large $V$, the site-centered CDW state is realized in order to
   avoid the increase in the energy of $V$.
As seen from Fig.~1, the site-centered CDW state 
  takes a maximum of the charge on the sites, 
  where the gain of the transfer energy is made favorable 
  owing to the electron hopping into both nearest sites.  
This state would correspond to the 4$\kf$ CDW-SP state 
 suggested by Clay \textit{et al.} (Fig.~3(d) in ref. \citen{Clay}) 
    and the CO+SP state suggested 
    by Kuwabara \textit{et al.},\cite{Kuwabara2003} 
    since the lattice tetramerization takes the maximum amplitude 
   on the sites. 
We note that, in the present case, 
the maximum of  charge is located on the same position as
  that of  lattice tetramerization (i.e., 
   the amplitude of $D_1$ is larger than that of $D_2$). 
 However,  for the state in ref. \citen{Kuwabara2003},    
    the maximum is obtained at the same position 
      as that of the $4\kf$ CDW (i.e., the amplitude of
 $D_2$ is larger than $D_1$).
 This may originate from our choice of relatively small $U$ 
 and then our site-centered CDW state 
 is expected to move to the CO+SP state  
      for large $U$.

\begin{figure}[t]
\centering
\includegraphics[width=8.5cm]{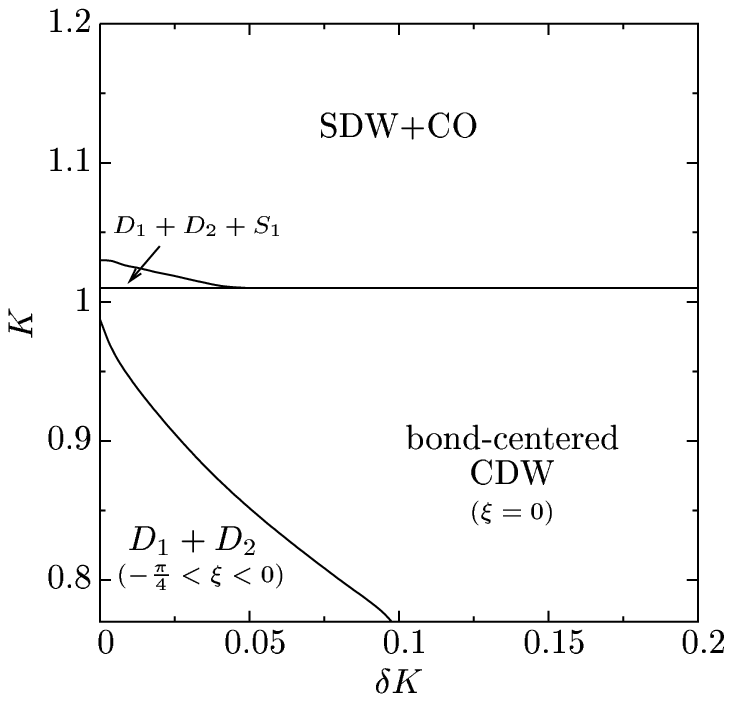}
\caption{
Phase diagram on the plane of $\delta K$ and $K$ for $U=3$ and $V=1$. 
The narrow region for small $\delta K$ and around $K\simeq 1.03$ is 
  characterized by $D_1+D_2+S_1$ (see Fig. 2).
With increasing $\delta K$, 
  the region $D_1$ (bond-centered CDW)  extends while 
 that for $D_1+D_2$ (site-centered CDW) diminishes.
For $\delta K \neq0$,  the $D_1 + D_2$ state takes $\xi$ with 
 an intermediate value of $ - \pi/4 <\xi <  0$.   
}
\end{figure}

  Now we show how the site-centered CDW ($D_1+D_2$) diminishes  when 
  $\delta K$ increases from zero. 
  In Fig.~3, the variations of the $D_1+D_2$ state 
   and the bond-centered CDW ($D_1$) are shown
  as a function of $\delta K$.
  The SDW+CO ($D_2+S_1$) remains unchanged since $\ut=0$. 
  The addition of the $\delta K$ term 
   increases the energy  of  the $D_1+D_2$ state 
   with $\xi \not= 0$, but does not affect on the $D_1$ state with 
   $\xi = 0$. 
   Thus, the change from  the $D_1+D_2$ state into the $D_1$ state 
 occurs with increasing $\delta K$.  
 In the phase diagram of Fig.~4,  
   the boundaries between  the $D_1+D_2$ state and 
      the $D_1$ state are shown for the choices 
       of $\delta K = 0.05$  and $0.1$.  
   The region of bond-centered CDW increases rapidly 
    with increasing $\delta K$, i.e.,   
 a small amount of  $\delta K$ 
  is enough to obtain the bond-centered CDW  state   as the ground state.
It is noticed that the boundary 
  of the direct transition from the 
   bond-centered CDW state to the SDW+CO state 
   emerges  even for $\delta K =0.05$. 
 Such a boundary is in contrast to that in Fig.~2.

\begin{figure}[t]
\centering
\includegraphics[width=8.5cm]{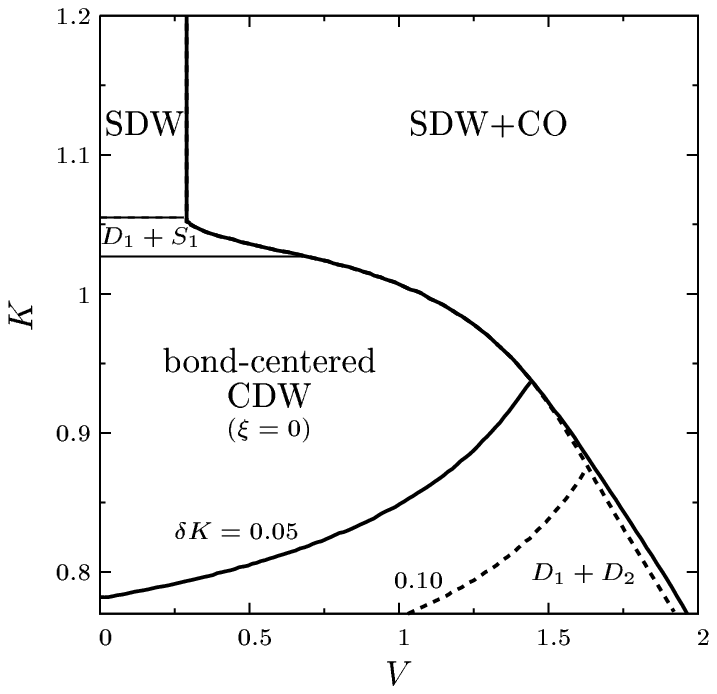}
\caption{
Phase diagram  on the plane of 
$V$ and $K$ with several choices of 
$\delta K = 0.05$  and 0.1. 
The phase boundaries 
  for finite $\delta K$
are shown by the dotted line ($\delta K=0.05$) 
 and the dashed line ($\delta K=0.10$).
The region of the bond-centered CDW ($D_1$) state is enlarged with 
increasing $\delta K$.  
For $\delta K \neq0$,  the state $D_1 + D_2$ takes $\xi$ with 
 an intermediate value of $ - \pi/4 <\xi <  0$.   
}
\end{figure}

Here, we compare  the experimental findings 
in the (EDO-TTF)$_2$PF$_6$ compound 
  with the results of 
the present calculation.
The Peierls state observed in (EDO-TTF)$_2$PF$_6$ is of the type of
  the bond-centered CDW, \cite{Drozdova2004}
 and  the photoinduced phase transition from 
 the Peierls insulator to a metal takes place by 
  the weak laser photoexcitation. 
This behavior has been discussed by assuming the metastable state
    without lattice distortion. \cite{Chollet2005}
In our calculation,
   the bond-centered CDW state is actually reproduced in Figs. 2 and 4.
One can expect the metastable state in the region  near the 
  boundary between the  the bond-centered CDW state 
   and  the undistorted state, since the  
   the phase transitions are of the first order.
Thus, we focus on the properties of the first-order phase transitions 
  from the bond-centered CDW state to the SDW+CO state.
From the inset of Fig. 2,
  it can be seen that the system does not show 
  a direct transition between  the bond-centered CDW state and  
  the  SDW+CO state, but there are  
         intermediate states  between these two states.
Actually, the following three types of transitions occur  
 when the bond-centered CDW state moves to the  SDW+CO state
  with increasing $K$. 
 (i) For $0.67 \lesssim V \lesssim 1.06$,
 the bond-centered CDW state shows a first-order phase transition 
     to the $D_1+D_2+S_1$ state 
     and undergoes  a second-order transition to the 
       SDW+CO state. 
(ii) 
For $0.49 \lesssim V \lesssim 0.67$, 
   the bond-centered CDW state shows a second-order transition 
  to the $D_1+S_1$ state followed by  the successive transitions of  
   a first-order one into the $D_1+D_2+S_1$ state and  
    a second-order one into the SDW+CO state. 
(iii) 
For $0.28 \lesssim V \lesssim 0.49$, 
   the bond-centered CDW state shows  a second-order transition 
     to the $D_1+S_1$ state 
     and then undergoes a first-order transition to the 
       SDW+CO state. 
 The first-order phase transition in  case (i)
   can be ascribed to the
   difference  between  the symmetry in the bond-centered CDW ($\xi=0$)
  and that in the $D_1+D_2+S_1$ state ($\xi=-\pi/4$), where both states 
   have  finite lattice distortion.
The origin of the first-order transition in  case (ii)  is
  also  the same as that in  case (i).
In these cases, 
 the second-order transition between the $D_1 + D_2 + S_2$ state 
 with $\xi=-\pi/4$  and the SDW + CO state implies that 
 the state around $\ut=0$ is  unstable  and turns out to be
   irrelevant  to  (EDO-TTF)$_2$PF$_6$.
For  case (iii), 
the first-order transition is due to the difference in 
 the  locking of the phase $\theta$,   i.e., 
   $\theta=\pi/4$ in the $D_1+S_1$ state and 
    $\theta=0$ in the SDW+CO state.
This case is also irrelevant to the state of 
 the (EDO-TTF)$_2$PF$_6$, since 
  the spin ordering is absent  in the Peierls state.
Thus, both the metastable state at $\ut=0$
    and the ground state of 
 the bond-centered CDW ($\ut \not= 0$) cannot be explained 
  within the model of $\delta K=0$,  suggesting a new mechanism for   
   the photoinduced phase observed in 
       the (EDO-TTF)$_2$PF$_6$ compound. 
 Our idea is the introduction of the alternation of the elastic constant,
 which enables us  to reproduce the experimental findings.
By increasing $\delta K$, the region of the bond-centered CDW state is 
  enhanced and moreover,  there is a direct first-order phase transition
  to the lattice-undistorted SDW+CO state, as seen from Fig.~4. 
 Such a behavior comes from the fact that the $\delta K$ has
  a role of fixing  the phase $\xi=0$, which 
    can be obtained from eq. (\ref{eq:MF_Hamiltonian}).
On the basis of such a consideration, 
 we examine the bond-centered CDW state and   
 the mechanism of the emergence of the metastable state at $\ut=0$
  by setting $\xi=0$ in the next section.  
We discuss  the relevance to the 
   (EDO-TTF)$_2$PF$_6$ compound  in \S 5.

\section{Phase Diagram and Metastable State for Large  $\delta K$}

In the following calculation,
 we examine the case of  $\xi = 0$
 corresponding to  the bond-centered  lattice distortion, 
   which is realized by choosing a moderate magnitude of 
   $\delta K \geq 0.2$ as shown in the preceding section.
The neglect of the lattice dimerization can be justified by 
  considering the large $\delta K$, as will be discussed in \S5.

\begin{figure}[b]
\centering
\includegraphics[width=8.5cm]{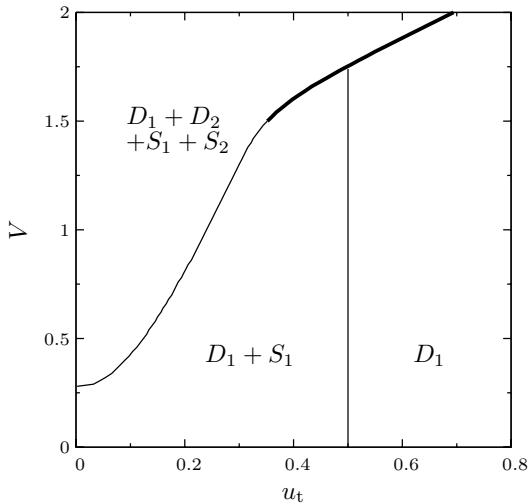}
\caption{
Phase diagram on the plane of  $\ut$ and  $V$ 
 where $\ut$ is regarded as an external field (see text).
 The thin and bold  lines correspond to the boundaries of
 the second-order and first-order transitions, respectively.  
Order parameters  $S_1, D_1, D_2$ represent 
     SDW, ECDW and CO, respectively.      
}
\end{figure}

In order to understand the role of 
  the  e-p interaction in the correlated system, 
       we first examine eq.~(\ref{eq:MF_Hamiltonian})
  by treating  $\ut$  as an external field, i.e., 
 by  discarding  the condition of eqs.~(\ref{eq:ut}) and (\ref{eq:xi}).
We calculate the order parameters $S_j$ and $D_j$ ($j=1$, $2$)  as 
   a function of $\ut$.
 The CDW, whose order parameter $D_1$
 is always  finite due to the presence of the 
  external field $\ut$,  
is called  the extrinsic 2$\kf$ charge-density wave 
   (ECDW). 
In Fig.~5, we explain how the state varies with increasing $\ut$.
 When  $V=0$, the order parameter $S_1$ 
  decreases monotonically and becomes zero at $\ut\approx 0.5$.
 For $V=1$,  both the order parameters $D_2$ and $S_1$ take
  finite values for $\ut=0$, but  as $\ut$ increases,
     $D_2$ decreases  rapidly and becomes zero at $\ut\approx 0.26$
     and $S_1$ becomes zero at $\ut\approx 0.5$.
    The $D_1+S_1$ state is   understood by  noting 
       that  $S_1$ is rather robust compared with $D_2$ because $U > V$.
With increasing  $\ut$ ($0.28 < V$),
the phase of the SDW, $\theta$,
     increases from zero and reaches $\pi/4$ at which $D_2$ vanishes. 
Such a variation of $\theta$ 
  comes from the competition  of $D_2$ and $D_1$.  
  We obtain  $\theta' = \theta + \pi/2$ 
    for the $D_1+S_1$ state and the $D_1$ state.    
For larger values of $V(\gtrsim 1.5)$, 
  the strong competition of $D_2$ and  $\ut$ results in  
  the first-order transition, which is followed by 
    the discontinuous changes of  $S_j$ and $D_j$ ($j=1$, $2$).
Furthermore, for  $V (\gtrsim 1.7)$,
  the vanishing of $D_2$ and $S_1$ takes place simultaneously
 due to the strong effect of $\ut$.

The phase diagram is summarized in Fig. 5.
 The second-order and first-order transitions take place 
    at the boundaries of
      the thin and  bold lines, respectively,
       where the point with  $\ut=0$ and $V=V_{\mathrm{c}}(\simeq 0.28)$
   is the singular one showing the first-order transition of $D_2$.  
For the coexistent state of $D_1+D_2+S_1+S_2$, 
   both the amplitudes $S_1$ and $D_2$ are larger than that of $D_1$.
 In the $D_1+S_1$ state, 
  the amplitude of $S_1$ is larger than that of $D_1$.
 For large $\ut$,  
  the pure ECDW ($D_1$) state  appears, 
      e.g., for $\ut \gtrsim 0.5$ and $V \lesssim 1.7$. 
The boundary between the $D_1+S_1$ state and 
  the $D_1+D_2+S_1+S_2$ state in the region of $0.5 \lesssim V \lesssim 1.5$  
is given  by  $ V \simeq 4 \ut$,
 indicating a direct competition between $\ut$ and $V$.
On the boundary corresponding to the vanishing of $D_2$, one sees  
   a point where
    the second-order transition changes into the first-order transition 
 with increasing $V$. 
This variation  resembles  that found in the phase diagram 
  in the $V$ versus temperature (instead of $\ut$) plane.
  \cite{Tomio2001JPCS}
 Such a fact  may be explained  
  by the existence of the term proportional to $D_2 S_1^2$  
  in the energy  expansion with respect to the order parameter.
  \cite{Tomio2001JPCS}
The solid line  of the phase boundary between the $D_1$ state
 and the $D_1+D_2+S_1+S_2$ state 
  is  given by $\ut \simeq  \frac{2}{3}V -0.8$  for  $V\gtrsim 1.7$. 
We found that $D_1$ is always finite  due to the external field $\ut$, 
 but is strongly suppressed in the presence of $D_2$ (not shown), i.e., 
 the strong competition between $D_2$ and $D_1$, 
 which is a characteristic of 1/4-filled systems. 
Also note that $\theta = \pi/4$ if $D_2 = 0$; i.e., 
  the maximum  density is located 
    on the bonds in the absence of CO.

Next we examine the $\ut$ dependence of the mean-field energy 
$E(\ut)$,  which is defined by $E(\ut,\xi=0)$.
The minimum of this energy gives the true ground-state energy.
In Fig.~6, the energy difference, 
  $\delta E(\ut) \equiv E(\ut) - E(0)$, 
is shown as a function of $\ut$ 
 for $V=0.1$ with some choices of $K$. 
This behavior denotes the conventional second-order phase transition
   except for its having  several kinds of order parameters, $S_1$ and $D_1$.
For $K =1.06$, the minimum is obtained at $\ut=0$ corresponding to 
  the pure SDW state. 
 The closed circle at $\ut \simeq  0.5$ denotes the point where 
     the $S_1$ state  vanishes. 
Thus, we obtain 
 the SDW ($S_1$) state for $K \gtrsim 1.05$,
 the $D_1+S_1$ state  for
 $1.03 \lesssim K \lesssim 1.05 $, 
  and
 the pure bond-centered CDW ($D_1$) state  for $K \lesssim 1.03$. 
Figure 7 shows the case for $V=1$.
The novel feature  in $\delta E(\ut)$ compared with that of Fig.~6 
 is the emergence of  a local minimum at $\ut=0$.
For  $0.79 \lesssim K \lesssim 1.01$, 
 we find a metastable state where 
 the energy of the local minimum $E(\ut=0)$ 
   is larger than the  energy of the true minimum 
  at finite $\ut$.
This fact implies that, with decreasing $K$,
   the system exhibits the first-order phase transition 
   from the SDW+CO ($D_2+S_1$) state into the bond-centered CDW 
   state  at $K \approx  1.01$.
We note that,  for $ 0.28 \lesssim V \lesssim 0.70$,
   there is a very narrow  region of $K$ 
   around $K \approx 1.03$  
   for the  $D_1+S_1$ state (not shown),  as found in Fig.~4.  
For a larger choice of $V = 1.6$, 
  the local minimum at $\ut=0$ is found for  $0.6 \lesssim K \lesssim 0.89$. 

\begin{figure}[t]
\centering
\includegraphics[width=6.5cm]{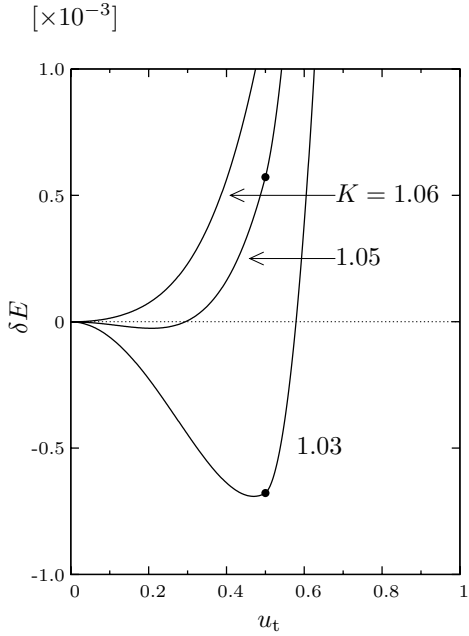}
\caption{
   $\ut$ dependence of $\delta E(\ut) (\equiv  E(\ut) - E(0))$ 
  for $V=0.1$ with fixed 
      $K=1.03$, $1.05$, and $1.06$.
  With increasing $\ut$,
        $S_1$ decreases  and vanishes at the point shown by 
  the closed circle. 
} 
\end{figure}
\begin{figure}[t]
\centering
\includegraphics[width=6.5cm]{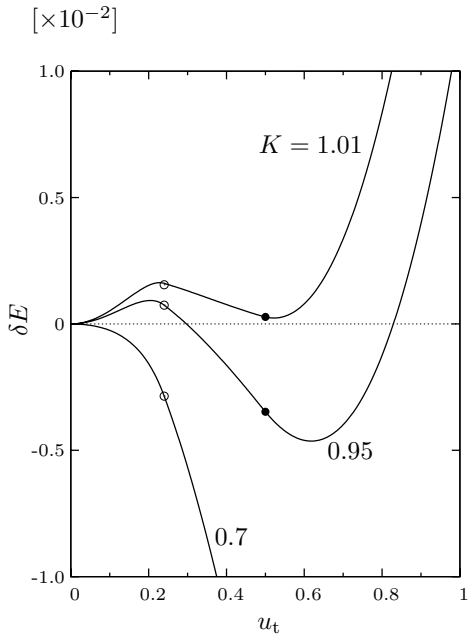}
\caption{
 $\ut$ dependence of $ \delta E (\ut)$ 
for $V=1$ with fixed 
 $K= 0.70$, $0.95$, and $1.01$.
 With increasing $\ut$,
  $D_2$ ($S_1$) decreases and vanishes  at the point shown 
  by the  open (closed) circle. 
}
\end{figure}

\begin{figure}[t]
\centering
\includegraphics[width=7cm]{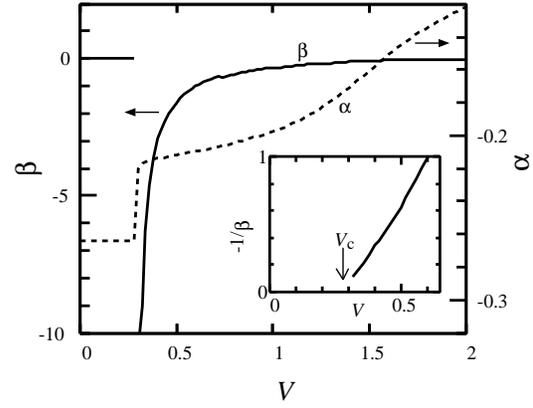}
\caption{
 $V$ dependence of  $\alpha$   and $\beta$ 
 in eq.~(\ref{eq:energy_expansion}), 
 where $\theta = \pi/4$ ($0$) for 
  $V < V_{\mathrm{c}}$ ($V>V_{\mathrm{c}}$). 
  The inset denotes $ - 1/\beta$, which 
   is proportional to $(V - V_{\mathrm{c}})$ for $V > V_{\mathrm{c}}$.   
}
\end{figure}

 Here, we examine the origin of the metastable state found in Fig.~7.
The mean-field energy of eq.~(\ref{eq:MF_Hamiltonian}) 
 can be expanded in terms 
  of $\ut$ as 
\begin{eqnarray}
\label{eq:energy_expansion}
  E(\ut) = E(0) + \left(\alpha + \frac{1}{4} K \right) \ut^2
   + \beta \ut^4 + \cdots .
\end{eqnarray} 
The coefficients, $\alpha$ and $\beta$, can be estimated numerically and 
   are shown in Fig. 8  as a function of $V$. 
    With increasing  $V$, 
  there is a jump in both  $\alpha$ and $\beta$ at 
  $V=V_{\mathrm{c}}(\simeq 0.28)$, 
   at which the pure SDW state moves to the SDW+CO state; 
    i.e., the  $D_2$ state emerges. 
It can be found that $\alpha < 0$ for arbitrary $V$, and that 
     $\beta < 0$ ($\beta > 0$) for $V>V_{\mathrm{c}}$ 
  ($V<V_{\mathrm{c}}$).   
The result of $\alpha < 0$ is quite reasonable when we note that
   $\ut$ acts as the external field. 
 In order to understand why  $\beta$ is negative and  
 diverges as $V \rightarrow V_{\mathrm{c}} + 0$, 
   we examine eq.~(\ref{eq:MF_Hamiltonian})
  in terms of the phase degrees of freedom $\theta$.
    \cite{Tomio2001JPCS}
 Actually, from eq.~(\ref{eq:MF_Hamiltonian}),
  the energy with fixed $\ut$ and $\theta$ is expressed as  
\begin{align}
\label{eq:energy_expansion2}
  E(\ut,\theta)  = &
   E_g(\theta)
 + \left( \alpha' - \gamma \sin 2 \theta   + \frac{1}{4}K \right)\ut^2 
  \nonumber \\ 
  & {} + \beta' \ut^4 + \cdots ,
\end{align}
where 
$E_g(\theta)$ is given by eq.~(\ref{eq:Eg}).
 Quantities $ \alpha'(<0)$, $\beta'(>0)$, and $\gamma (>0)$ are estimated  
  numerically  from eq.~(\ref{eq:MF_Hamiltonian}).
   The  $\gamma$-term, which is proportional to 
    $S_1^2$, denotes the interaction 
  between the bond-centered CDW ($D_1$) state 
       and the SDW ($S_1$) state.
The coefficients are estimated as  
 $ \alpha' \simeq -0.197$, $\beta'=0.0068$,
  and  $\gamma \simeq 0.056$ for $V=1$,
where  $\beta' \simeq 0.0145$ for $V < V_{\mathrm{c}}$.  
  There is a small jump of $\gamma$ at $V = V_{\mathrm{c}}$, 
    where the $V$ dependence of $\gamma$ is much smaller  than 
     that of $\alpha.$ 
The locking position of $\theta$ can be determined in order to 
  minimize  eq.~(\ref{eq:energy_expansion2}).
The commensurability potential $E_g(\theta)$ 
  for $V>V_{\mathrm{c}}$ favors the locking position  $\theta = 0$,
  while the $\gamma$-term favors $\theta=\pi/4$.
For $V>V_{\mathrm{c}}$, 
by minimizing eq. (\ref{eq:energy_expansion2}) with respect to $\theta$,
   we obtain eq.~(\ref{eq:energy_expansion}) with the coefficients 
 \begin{align}
\alpha = \alpha' ,
 \qquad
\beta  = \beta' - \frac{\gamma^2}{8 |C(V)|} .
\label{eq:energy_expansion3}
\end{align}
We can immediately reproduce the anomalous behavior 
   $-1/\beta \propto (V-V_{\mathrm{c}})$ at $V=V_{\mathrm{c}}+0$,
  by noting that the coefficient $C(V)$ follows
   $C(V)\propto (V-V_{\mathrm{c}})$.
This is the reason why the coefficient of the  fourth-order term $\beta$
   can become negative for $V>V_{\mathrm{c}}$.
It is shown that the result similar to  eq.~(\ref{eq:energy_expansion3}) 
can be obtained from the bosonization scheme, in which 
the spin degree of freedom is also taken into account.
 \cite{Tsuchiizu2007_preprint}  
 For $V< V_{\mathrm{c}}$, 
 the $\gamma$-term does not contribute to the $\ut^4$ term 
  since both the commensurability potential and 
   the $\gamma$ term favors the locking position $\theta =  \pi/4$
  (one simply obtains $\alpha=\alpha'-\gamma$).
Thus, the conventional second-order transition is reproduced.
From these arguments,
  the conditions for the metastable state are summarized as follows:
{\it 
(i) the existence of the CO  for $V>V_{\mathrm{c}}$ 
  (the commensurability energy leading to $\theta=0$),
(ii) the coupling between the SDW state ($S_1$) and 
 the bond-centered CDW state ($\ut$), which is  given by  the $\gamma$ term 
 in eq.~(\ref{eq:energy_expansion2}), and  
(iii) the mechanism to fix $\xi=0$ as discussed in \S3. 
 }

\begin{figure}[t]
\centering
\includegraphics[width=7.5cm]{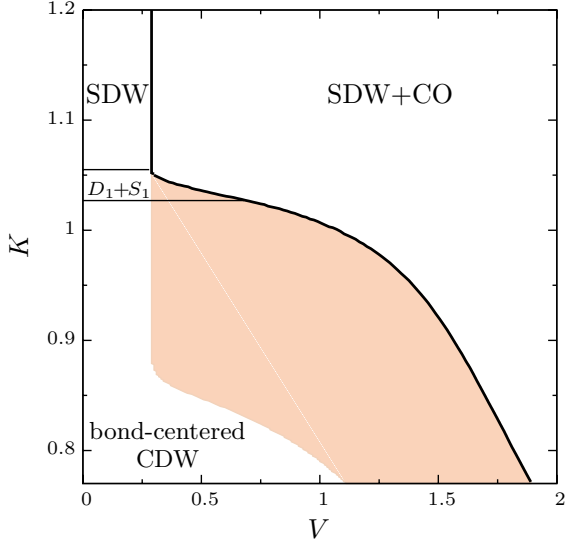}
\caption{ (Color online)
 Phase diagram of  the
 bond-centered CDW ($D_1$), SDW+CO ($D_2 + S_1$), SDW  ($S_1$), and
  bond-centered-CDW+SDW ($D_1+S_1$)  states,
  on the plane of $V$ and $K$.
  The shaded region denotes the area where 
    the metastable state at  $\ut = 0$ appears.
  The thin and bold lines correspond  
   to phase boundaries of the second-order and first-order transitions,
   respectively.
}
\end{figure}

 Finally we show the phase diagram of the ground state on the plane of 
     $V$ and $K$ in Fig.~9. 
 For $K \gtrsim 1.05$ corresponding to the weak e-p coupling,  
  we obtain either the pure SDW ($V<V_{\mathrm{c}}$) 
   with $\theta =  \pi/4$  or 
  the SDW+CO ($V>V_{\mathrm{c}}$) with $\theta = 0$.
 With decreasing  $K$ (i.e., increasing the e-p coupling), 
  the bond-centered CDW state with $\theta=  \pi/4$  
     appears for $K <K_{\mathrm{c}}$ where $K_{\mathrm{c}}$
   is a critical value of the first-order phase transition 
   for the Peierls state. 
The state with $\ut=0$ becomes metastable 
    in the large interval region of $K < K_{\mathrm{c}}$ 
     as shown by the shaded area. 
   The lower boundary of the shaded region is given by 
 \begin{eqnarray}
   K = - 4 \alpha ,
 \end{eqnarray}  
 which is verified from $\alpha$ in Fig.~8. 
 There is also another metastable state 
  with the local minimum at $\ut \not=0$,
  which is  located in a certain region of $K > K_{\mathrm{c}}$
  (not shown in Fig. 9).

\section{Summary and Discussion}

We examined the Peierls-Hubbard model at quarter-filling 
 with the intersite interaction ($V$), and 
 obtained the ground-state phase diagram within the mean-field theory.
 A noticeable finding is the undistorted state ($\ut=0$), 
 which becomes metastable in the bond-centered CDW 
  state close to the boundary  between 
  the bond-centered CDW state and the SDW+CO state.
  The obtained bond-centered CDW ground state and its metastable
  state may share the following  common features 
   with the Peierls state of the (EDO-TTF)$_2$PF$_6$ compound.
   The spatial variation of the  Peierls state is the same, 
    and  the metastable state for $\ut=0$ 
  could be related to the photoinduced phase found in the experiment 
     on the EDO-TTF compound.

  In the present analysis of \S4,  we assumed  $\xi = 0$
    in order to stabilize  the bond-centered CDW state
  and examined the metastable state at $\ut=0$,
 which may be relevant to the property of  the EDO-TTF compound.
 Actually,  in the normal state of this compound, there is the alternation of 
   the bending of the molecule for every two sites along the   
   the one-dimensional chain.\cite{Ota2002}
In addition to the ground state (\S 3), 
 we discuss this assumption for the metastable state 
      on the basis of the energy, which is  expanded in terms of $\ut$.
 The energy  with fixed $\ut$, $\theta$, and $\xi$ 
 can be explicitly written  as 
\begin{align}
\label{eq:xi2}
E(\ut, \theta,\xi) 
=& \,
  C(V) \cos 4 \theta 
   + \left( \alpha'  - \gamma \sin (2 \theta - 2 \xi)  \right)  \ut^2 
     \nonumber  \\  
 &+  \left( \frac{K}{2} + \frac{\delta K}{4}   \sin^2 \xi \right) \ut^2  
 +  \cdots  ,   
\end{align}
where the $\delta K$  term denotes an increase in elastic energy 
 for $\xi \not= 0$.
Here, we note that, even for $V>V_{\mathrm{c}}$,
  there is no competition between $\theta$ and $\xi$ for $\delta K=0$,
   since  the minimum of eq.~(\ref{eq:xi2})
    is obtained by choosing $\xi$  so as to satisfy 
   $\sin (2 \theta - 2 \xi) = 1$. 
 In this case, the  $\gamma^2/C(V)$ term in 
 $\beta$ [see  eq.~(\ref{eq:energy_expansion3})] is absent and
the energy $E(\ut, \theta,\xi)$ [eq. (\ref{eq:xi2})]
 takes a true  minimum  at $\ut=0$ (i.e., $\beta > 0$).
When  $\delta K \neq 0$, 
  the energy $E(\ut, \theta,\xi)$  as a function of $\xi$ 
  increases and the
  metastable state around $\ut=0$ can be expected (i.e., $\beta < 0$).
Actually, for $\delta K>0.46$, the energy $E(\ut, \theta,\xi)$
   takes a local minimum at $\ut=0$ for all values of $\xi$, 
  in the case of $K=0.9$ and $V=1$.
  Thus,  the  assumption $\xi=0$ can be  justified when  
   the  degree of alternation of the elastic constant 
    $\delta K$ becomes   larger than a critical value. 

In the present analysis, we did  not consider  the
  possibility of lattice dimerization
  ($u_j=(-1)^j u_{\mathrm{d}}$, where
  $u_{\mathrm{d}}$ is the amplitude of the lattice dimerization).
Since  the elastic energy of the lattice dimerization is given by
  $(N/2)(K+\delta K/2) u_{\mathrm{d}}^2 $,  
  $\delta K$ also has an effect of suppressing 
   the lattice dimerization. 
From these arguments, we expect that the 
 effect of $\delta K$ will arise from the bending freedom of the molecules 
     in  the (EDO-TTF)$_2$PF$_6$ compound.  
Such a bending plays important roles in the photoinduced cooperative 
  transition,
    although  the origin of the molecular property   
    still remains an open question 
    from a microscopic view point.

 Here, we note states at finite temperatures. 
There are some theoretical works 
 on  finite-temperature properties.
Quite recently, it has been pointed out that 
the  Peierls-Hubbard model with $\delta K = 0$ exhibits   the first-order  
transition 
  from the dimer-Mott state into the spin-Peierls state,\cite{Seo2007JPSJ}  
where both states show the insulating behavior.
On the other hand,
the purely electronic model shows the transition  
from the metallic state 
     into the insulating state 
of the 2$\kf$ SDW + 4$\kf$ SDW + 2$\kf$ CDW state.
          \cite{Tomio2001JPCS}  
 Compared with the state of the (EDO-TTF)$_2$PF$_6$ compound, 
   the normal state at high temperatures is different from 
  the dimer-Mott state obtained from the former model  and 
    the lattice deformation cannot be reproduced in the latter model. 
It is expected that the parameters 
  in the shaded region in Fig. 9
  will lead the first-order phase transition into the normal state 
    with increasing temperature, 
    and that 
 such a transition will be strongly enhanced     
by  the effect of the large bending 
  of molecules in the (EDO-TTF)$_2$PF$_6$.

\acknowledgements

The authors thank  Professors   H. Yamochi, G. Saito,  S. Koshihara, 
K. Yonemitsu, 
J.-F. Halet, and L. Ouahab,  and 
  Drs. Y. Nakano, K. Onda, A. Ota, and H. Seo,
 for useful discussions and comments. 
     The present work has been financially supported
 by a Grant-in-Aid for Scientific Research on Priority Areas
      of Molecular Conductors (No. 15073213)
 from the Ministry of Education, Culture, Sports, 
Science and Technology, Japan, 
and by
the JSPS Core-to-Core Program Project 
^^ ^^ Multifunctional Molecular Materials and
Device Applications".

\end{document}